\documentclass[eps,preprint,12pt]{revtex4}

\usepackage{amsmath}
\usepackage{amssymb}
\usepackage{amsfonts}
\usepackage{graphicx}
\usepackage{dcolumn}
\usepackage{bm}
\usepackage{hyperref}
\hypersetup{colorlinks=false}

\begin{document}

\title{Casimir effect due to a slowly rotating source in the weak field approximation}

\author{V. B. Bezerra}
\address{Departamento de F\'isica, Universidade Federal da Para\'iba \\
58.059-970, Caixa Postal 5.008, \\ Jo\~ao Pessoa, PB, Brazil.}

\author{H. F.
Mota}
\address{Departamento de F\'isica, Universidade Federal da Para\'iba \\
58.059-970, Caixa Postal 5.009, \\ Jo\~ao Pessoa, PB, Brazil.}

\author{C. R. Muniz}

\address{Grupo de F\'isica Te\'orica(GFT), Universidade
Estadual do Cear\'a, UECE-FECLI\\
 Iguatu, Cear\'a, Brazil.}

\date{\today}

\begin{abstract}

We calculate the renormalized vacuum energy density for a massless scalar field confined between two nearby parallel plates formed by ideal uncharged conductors, placed very close to the surface of a rotating spherical gravitational source with mass $M$, radius $R$ and momentum angular $J$, at the equatorial region. We consider that the source rotates slowly and that the gravitational field is weak. Corrections to the Casimir energy density induced by the gravitational field generated by this source are calculated up to $M/R^2$ order. The obtained results show us that there is an important modification in the Casimir energy only in this order of approximation, which depends on the surface gravity as well as on the rotation of the source. Thermal corrections to the Casimir energy found also are calculated in all these orders.

\vspace{0.75cm}
\end{abstract}

\maketitle

\section{Introduction}

The Casimir effect is a phenomenon which appears in different areas of physics and has numerous applications in condensed matter, atomic physics, elementary particles and gravitation and cosmology, among others. It was predicted by H. B. G. Casimir in the later 1940´s as an attractive force between two parallel and uncharged metallic plates placed in vacuum \cite{casimir1}. At the classical level, there is no force between two uncharged conductor plates, thus, the Casimir effect is of quantum origin and results from the modifications of the zero-point(vacuum) oscillations of the electromagnetic field induced by the material boundaries compared to the free Minkowski spacetime.

The first experimental attempt to confirm the Casimir force was made by Sparnaay \cite{sparnaay}. Since then, a lot of experiments were performed on measuring the Casimir force between conductor, semiconductor and dieletric bodies (see the reviews \cite{klim}-\cite{capasso}). More recently, important experiments were performed \cite{banishev1}-\cite{banishev5}. On the other hand, from the theoretical point of view, a considerable number of theoretical works presents generalizations of the pioneering formulation by Casimir, and these includes various geometrical configurations of the plates made of different materials at any non-zero temperature, quantum fields of different spins, space-times with different geometries and topologies \cite{mostepanenko}, and, finally, different gravity theories \cite{petrov}-\cite{celio}.

The Casimir effect in gravitational fields with non-trivial topology was studied since the middle 1970´s in the framework of quantum field theory in curved space-times \cite{dewitt}, and in particular in an Einstein universe with a topology $R^{1} \times S^{3}$, for a massless scalar field \cite{larry}. In such context, there are no material boundaries, but there are some identification conditions imposed by topology to the quantum fields which play the same role as the boundary conditions associated with material boundaries. Along this line, a considerable number of works was devoted to the Casimir effect in space-times with non-Euclidean topology \cite{dowker1}-\cite{eugenio2}.

Recently, the vacuum quantum effect on a massless scalar field confined between two material boundaries, in a static spacetime, in the weak-field approximation, was considered by Sorge \cite{sorge}. In that work, nearly parallel plates formed by ideal conductors were placed at surface of a sphere of mass $M$ and radius $R$, and corrections to the Casimir energy as compared with the one in Minkowski spacetime, induced by gravity, up to second order in $(M/R)$, were found. The same Casimir apparatus, but now taking into account the effects of rotation due to a Kerr black hole, in the weak field approximation, was analysed by Zonoz and Nazari \cite{zonoz}. In this last paper, a different approach was used on the basis of an effective refraction index associated to the spacetime, and it was shown that the correction to Minkowski Casimir energy of the electromagnetic field, induced by gravity, depends on the spacetime rotation.

Following the approach developed by Sorge \cite{sorge}, we calculate the corrections to the flat Casimir energy of a massless scalar field confined in a cavity formed by two parallel conducting and uncharged plates. The distance between them, $L$, is such that $L\ll R$, where $R$ is the radius of the source. The plates are placed tangentially to the equator line of the rotating sphere with angular momentum $J$ and mass $M$. The source rotates slowly and the weak field regime is taken into account. We also will calculate the thermal corrections to the Casimir energy from the renormalized Helmholtz free energy, as well as the internal energy in both low and high temperature limits. It is worth calling attention to the fact that the Casimir apparatus considered as well as the source of the gravitational field show us what is the role played by the rotation and temperature if we compare the obtained results with the ones found by Sorge \cite{sorge}.

The paper is organized as follows. In section 2, we calculate the corrections to the Casimir energy up to $M/R^2$ order of approximation, in spacetime generated by a slowly rotating source, at $T=0$. In section 3, we find the thermal corrections. In section 4, we present the final remarks.

\section{Gravitational corrections to the Casimir energy at zero temperature}

Firstly, let us consider the spacetime metric corresponding to a slowly rotating source in the weak field approximation. It is given, in isotropic coordinates, by the following expression

\begin{equation}\label{01}
ds^2=[1+2\phi(r)]dt^2-[1-2\phi(r)](dr^2+r^2d\theta^2+r^2\sin^2{\theta}d\varphi^2)+4a \sin^2{\theta}\phi(r) dt d\varphi,
\end{equation}
In fact, this metric corresponds to the Kerr metric considering only linear terms in the angular momentum and taking into account the weak field approximation regime. The function $\phi(r)=-M/r$ is the post-Newtonian parameter and $a=J/M$ is the angular momentum per mass parameter. We will work with the plates positioned on the equatorial region of the rotating sphere, and thus $\theta=\pi/2$, exactly where the searched effects due to rotation are more pronounced.

The isotropic coordinates are introduced in order to consider the geometry of the cavity, enabling the transformation in spatial line element $dr^2+r^2d\theta^2+r^2\sin^2{\theta}d\varphi^2\rightarrow dx^2+dy^2+dz^2$, and such that the distance between the plates is parallel to the $x$ axis.

The unities in which $\hbar=c=G=1$ will be used here.

\subsection{First order correction to the Casimir energy}

Let us consider that the post-Newtonian parameter due to the massive sphere in the region between the plates can be expanded as
\begin{equation}\label{02}
\phi(r)=-\frac{M}{R+x}\simeq\Phi_0(R)+\gamma x,
\end{equation}
where $\Phi_0(R)=-M/R$, $\gamma=M/R^2$ and $x$ is the coordinate of a point inside the cavity if the origin of the reference frame is on one of the plates. In this section, we consider $\gamma=0$. Thus,
the metric (\ref{01}), turns into
\begin{equation}\label{03}
ds^2\simeq(1+2\Phi_0)dt^2-(1-2\Phi_0)(dx^2+dy^2+dz^2)+4a\Phi_0dtd\varphi,
\end{equation}
Now, let us introduce a rotating reference frame with respect to which the plates stay at rest. This new Cartesian reference frame, whose $z'$ axis coincides with the rotation axis of the source, $z$, is related to that introduced in Eq. (\ref{03}), after one moving its origin to the center of the sphere, in such manner that $(x',y',z')=(x\cos{\Omega t}+y\sin{\Omega t},-x\sin{\Omega t}+y\cos{\Omega t}, z)$, following \cite{konno}. This turns the spatial line element of Eq. (\ref{03}) invariant with respect to the rotation. The azimuthal angle must vary, therefore, as $\varphi=-\Omega t$, where $\Omega$ is the angular velocity of the reference frame, which is assumed to be the same one of the gravitational source, at least on the equator line. Of course, the angular velocity $\Omega$ is related to the parameter $a$, that depends on the angular momentum of the source. Note that also there is a natural limitation on the radial distances in this case, which is given by the condition $r<c/\Omega$ \cite{landau}. For the Earth, with $R\simeq6\times10^6$ m, $c/\Omega\sim10^{11}$ m and, therefore, we have no trouble with the rotating reference frame at its surface. It is important mentioning the works by Letaw and Pfautsch \cite{letaw1}-\cite{letaw3}, in which is shown that the rotating coordinate system defined on the whole spacetime is not necessarily stationary. We consider that, due to the present study to be limited to a very small region of spacetime, the introduction of such rotating referential frame is permitted. Unruh-like back reaction effects on the plates also will be neglected \cite{suga}, due to low rotation of the system.

Thus, the metric (\ref{03}) can be written in the diagonal form as
\begin{equation}\label{04}
ds^2\simeq(1+2b\Phi_0)dt^2-(1-2\Phi_0)(dx^2+dy^2+dz^2),
\end{equation}
where $b=1-2a\Omega$.
Now, let us solve the Klein-Gordon equation for the massless scalar field in vacuum, in the spacetime given by Eq. (\ref{04}), which is given by

\begin{equation}\label{05}
\frac{1}{\sqrt{-g}}\partial_{\mu}(\sqrt{-g}g^{\mu\nu}\partial_{\nu})\Psi=[1-2(b+1)\Phi_0]\partial_t^2\Psi-\nabla ^{2}\Psi=0.
\end{equation}

Let us assume that the solution of Eq. (\ref{05}) can be written as
\begin{equation}\label{06}
\Psi_{n,1}(x,y,z,t)=N_{n,1}\exp{[i( k_yy+k_zz-\omega_{n,1} t)]}\sin{(n\pi x/L)}
\end{equation}
such that the Dirichlet boundary conditions are satisfied on the plates. Thus, doing the substitution of Eq. (\ref{06}) into Eq. (\ref{05}), we find the eigenfrequencies $\omega_{n,1}$. Taking into account the approximation up to $(M/R)$ order we get
\begin{equation}\label{07}
\omega_{n,1}=[1+\Phi_0(b+1)]\left(k_y^2+k_z^2+\frac{n^2\pi^2}{L^2}\right)^{1/2}.
\end{equation}

In order to set the vacuum expectation value of temporal component of the energy-momentum density tensor, we need to perform the calculation of the normalization constant $N_n$, from the norm of the scalar functions $\Psi_n$ (which obey the usual orthonormality conditions) defined on the spacelike $\Sigma$ Cauchy hypersurface as \cite{birrel},
\begin{equation}\label{08}
\|\Psi_n\|=i\int_{\Sigma}\sqrt{-g_{\Sigma}}[\partial_t(\Psi_n)\Psi_n^{*}-\partial_t(\Psi_n^{*})\Psi_n]n^{t}d\Sigma,
\end{equation}
where $g_{\Sigma}$ is the determinant of the induced metric on the hypersurface $g_{ik}$, $i,k = 1,2,3$ and $d\Sigma=dxdydz$. In Eq. (\ref{08}), we have defined a timelike future-directed unitary quadrivector ${\bf n^t}$ as
\begin{equation}\label{09}
{\bf n^t}\equiv (n^t,0,0,0)=(1-b\Phi_0,0,0,0),
\end{equation}
and, thus, we find that the square of the normalization constant is
\begin{equation}\label{10}
N^2_{n,1}=\frac{1}{(2\pi)^2[1-(b+3)\Phi_0]L\omega_{n,1}}.
\end{equation}

The local vacuum expectation value of the scalar field energy density $T_{tt}(\Psi_n)$ is given by
\begin{equation}\label{11}
\epsilon_{vac,1}=<0|n^tn^tT_{tt}(\Psi_{n})|0>=\int d^2k_{||}\sum_nn^tn^tT_{tt}(\Psi_n),
\end{equation}
where $d^2k_{||}=dk_ydk_z$ and
\begin{equation}\label{12}
T_{tt}(\Psi_n)=\frac{1}{2}\left(\partial_t\Psi_n\partial_t\Psi^{*}_n-g_{tt}g^{ik}\partial_i\Psi_n\partial_k\Psi^{*}_n\right).
\end{equation}
The vacuum energy density $\overline{\epsilon}_{vac}$ is obtained by averaging $\epsilon_{vac}$ in the spatial
region (cavity) of the Casimir aparatus, whose expression is
\begin{equation}\label{13}
\overline{\epsilon}_{vac}=\frac{1}{V_p}\int_{\Sigma}\sqrt{-g_{\Sigma}}\epsilon_{vac}d\Sigma
\end{equation}
where $V_p=\int_{\Sigma}\sqrt{-g_{\Sigma}}d\Sigma$ is the proper volume of the cavity.
Putting (\ref{12}) into (\ref{11}) and this into (\ref{13}), after solving the spatial integrals we find
\begin{eqnarray}\label{14}
\overline{\epsilon}_{vac}=\left(1+4\Phi_0\right)\int\frac{d^2k_{||}}{2(2\pi)^2L}\sum_n\left[k^2_{||}+\left(\frac{n\pi}{L}\right)^2\right].\nonumber\\
\end{eqnarray}
The integral in the rhs of Eq. (\ref{14}) is just the vacuum energy density of the massless scalar field in Minkowski space, which diverges. The Casimir energy density $\overline{\epsilon}^{ren}_{vac}$ is found by using an adequate
renormalization procedure, as the Abel-Plana subtraction formula \cite{bordag}, for example. Applying this renormalization process in Eq.(\ref{14}), we get
 \begin{equation}\label{15}
\overline{\epsilon}^{ren}_{vac}=-(1+4\Phi_0)\frac{\pi^2}{1440L^4}.
\end{equation}

Taking into account the proper length $L_p=\sqrt{g_{xx}}L\simeq(1-\Phi_0)L$, the Casimir energy given by Eq. (\ref{15}), in terms of this length $L_p$, becomes
\begin{eqnarray}\label{16}
\overline{\epsilon}^{ren}_{vac}=-\frac{\pi^2}{1440L_p^4}.
\end{eqnarray}
This means that no gravitational or rotational effects on Minkowsky Casimir energy appear in $M/R$ order of approximation.

\subsection{Second order correction to the Casimir energy}

In what follows, we will consider second order corrections, in terms of $(M/R)$, to the Casimir energy and in low-rotation Kerr spacetime. As the Casimir effect is not affected in first order approximation of $M/R$, we will take $\Phi_0=0$. The metric to be used here is
\begin{equation}\label{17}
ds^2\simeq(1+2\gamma x)dt^2-(1-2\gamma x)(dx^2+dy^2+dz^2)+4\gamma x a d\varphi dt,
\end{equation}
where the origin of the system of axis is now on one of the plates. Passing to the rotating reference frame, moving with angular velocity $\Omega$ around $z$ axis, we have
\begin{equation}\label{18}
ds^2\simeq(1+2b\gamma x)dt^2-(1-2\gamma x)(dx^2+dy^2+dz^2),
\end{equation}
where $b=1-2a\Omega$, as in the previous subsection and $\gamma x$ is of $\mathcal{O}(M/R)^{2}$.

The Klein-Gordon equation in the spacetime with the metric given by (\ref{18}) is
\begin{equation}\label{19}
[1-2(b+1)\gamma x]\partial_t^2\Psi-\nabla^{2}\Psi=0.
\end{equation}

Let us seek for solutions of the equation (\ref{19}) of the form
\begin{equation}\label{20}
\Psi_{n,2}(x,y,z,t)=N_n\exp{[i(k_yy+k_zz-\omega t)]}\chi (x),
\end{equation}
where $\chi(x)$ is a function to be determined. Substituting Eq. (\ref{20}) into Eq. (\ref{19}), we get the following equation for $\chi(x)$
\begin{equation}\label{21}
\frac{d^2\chi}{dx^2}-4\widetilde{\gamma}\omega^2(b+1)x\chi+(\omega^2-k_{||}^2)\chi=0
\end{equation}
where
\begin{equation}\label{22}
\widetilde{\gamma}=\frac{\gamma(b+1)}{2}
\end{equation}
The solution $\chi(x)$ is given in terms of the Airy function $\text{Ai(z(\textit{x}))}$ in form
\begin{equation}\label{23}
\chi(x)\propto\text{Ai}\left[\frac{k_{||}^2-\omega^2+4 \widetilde{\gamma} \omega^2 x}{2\ 2^{1/3} \left(\widetilde{\gamma} \omega^2\right)^{2/3}}\right].
\end{equation}
Using the asymptotic expansion for large arguments of Airy function \cite{abramowitz}
\begin{equation}\label{24}
\text{Ai($-$z)}\sim \frac{\sin{(\frac{2}{3}\text{z}^{3/2}+\frac{\pi}{4})}}{\sqrt{\pi}\text{z}^{1/4}},
\end{equation}
and considering the Dirichlet boundary conditions, we find the following eigenfrequencies
 \begin{equation}\label{25}
 \omega_{n,2}^2=(1+2\widetilde{\gamma} L)\left(k_{||}^2+\frac{n^2\pi^2}{L^2}\right)=[1+\gamma(b+1)L]\left(k_{||}^2+\frac{n^2\pi^2}{L^2}\right),
 \end{equation}
where, in second equality, we have used equation (\ref{22}).

In order to calculate the quantum vacuum energy for this new order of approximation, let us assume that the solution of equation (\ref{19}) can also be written as a correction of the harmonic modes obtained in the case when the post-Newtonian parameter is expanded up to $\mathcal{O}(M/R)$, presented in the previous subsection. Let us represent the solution for that order as $\Psi_{n,2}$. Thus, we can write
\begin{equation}\label{26}
\Psi_{n,2}=\Psi_{n,1}+\delta\Psi_{n,1},
\end{equation}
where $\Psi_{n,1}$ is given by equation (\ref{06}). Therefore, the expression for the quantum vacuum energy can be separated as
\begin{equation}\label{27}
\overline{\epsilon}_{vac,2}=\overline{\epsilon}_{vac}+\delta\overline{\epsilon}_{vac}
\end{equation}
where
\begin{equation}\label{28}
\overline{\epsilon}_{vac}=\frac{1}{V_p}\int d^2k_{||}\sum_n\int_{\Sigma}\sqrt{-g_{\Sigma}}n^tn^t T_{tt}(\Psi_{1,n})d\Sigma.
\end{equation}
Employing the same procedure used in previous subsection to calculate the quantum vacuum energy and taking into account that, in present situation, $n^t\simeq1-b\gamma x$, we get
\begin{equation}\label{29}
\overline{\epsilon}_{vac}=\left[1+(b-1)\gamma L\right]\int \frac{1}{2(2\pi)^2L} d^2k_{||}
\sum_n \omega_{0,n},
\end{equation}
where $\omega_{0,n}=\left(k^2_{||}+\frac{n^2\pi^2}{L^2}\right)$. The renormalization of the integral which appears in Eq. (\ref{29}) results in the flat Casimir energy density, and then
 \begin{equation}\label{30}
\overline{\epsilon}^{ren}_{vac}=-\left[1+(b-1)\gamma L\right]\frac{\pi^2}{1440 L^4}.
 \end{equation}
Now, our task is to calculate $\delta\overline{\epsilon}_{vac}$, according to \cite{sorge}
 \begin{equation} \label{31}
 \delta\overline{\epsilon}_{vac}=\frac{1}{2(2\pi)^2L}\int d^2k_{||}\sum_n(\omega_{2,n}-\omega_{0,n})=\frac{(b+1)\gamma L}{2}\frac{1}{2(2\pi)^2L}\int d^2k_{||}\sum_n\omega_{n,0}.
 \end{equation}
 Finally, the second order correction to the renormalized Casimir energy, taking into account the proper length, given by $L\simeq[1+(1/2)\gamma L_p]L_p$, is

 \begin{equation}\label{32}
 \overline{\epsilon}^{ren}_{vac,2}=\overline{\epsilon}^{ren}_{vac}+\delta\overline{\epsilon}^{ren}_{vac}=-\left[1-{\gamma L_p(1 + 3a\Omega)}\right]\frac{\pi^2}{1440L_p^4}.
 \end{equation}

This result can be understood as the Casimir energy density for the flat spacetime with a correction term that depends on the surface gravity, rotation of the spacetime and non-inertial effects due to the rotating reference frame, codified in $\gamma$, $a$ and $\Omega$, respectively. It is worth noting that when $b=1$ ($a,\Omega=0$), we get the same result obtained by Sorge \cite{sorge}.

\section{Thermal corrections}

Let us investigate now what is the role of the temperature on the Casimir energy. We aim to calculate corrections to the latter due to a thermal bath at absolute temperature $T$. To do this, we need to find the renormalized Helmholtz free energy $\Delta_T\mathcal{F}_0^{ren}$, which is given by \cite{svaiter,bordag}
\begin{equation}\label{33}
\Delta_T\mathcal{F}_0^{ren}=2Ak_BT\sum_{n=0}^{\infty}\int\int\frac{d^2\textbf{k}_{||}}{(2\pi)^3}\log{\left[1-e^{-\widetilde{\beta}\left(\frac{n^2\pi^2}{L^2}+k^2_{||}\right)^{1/2}}\right]}-V_pf_{bb}(T),
\end{equation}
where $A$ is the area of the plates, $k_B$ is the Boltzmann constant, $V_p$ is the proper volume and $\widetilde{\beta}=F/k_B T$. This factor $F$ will be given by $F^{(1)}=[1+\Phi_0(b+1)]$, if we are performing calculations in $M/R$ order of approximation, or $F^{(2)}=[1+\gamma(b+1)L]$, if terms $\mathcal{(M/R)^2}$ are taken into account.  The second term in the rhs of Eq. (\ref{33}) is the free energy density of black-body radiation, expressed by
\begin{equation}\label{34}
f_{bb}(T)=k_BT\int\int\int\frac{d^3\textbf{k}}{(2\pi)^3}\log{\left[1-e^{-\widetilde{\beta}\left(k_x^2+k^2_{||}\right)^{1/2}}\right]},
\end{equation}
which is the non-confined part of the free energy density.

\subsection{First order correction to the Casimir energy}

The black-body free energy density (\ref{34}) can be easily found, in the approximation used here, as being given by
\begin{equation}\label{35}
f^{(1)}_{bb}(T)=-\left(1+\frac{6M}{R}-\frac{6Ma\Omega}{R}\right)\frac{\pi^2(k_BT)^4}{90}.
\end{equation}
We remark that, in the absence of gravity, Eq. (\ref{35}) is the flat black-body free energy density.

Expanding the logarithm in Eq. (\ref{33}), we get
\begin{equation} \label{36}
\Delta^{(1)}_T\mathcal{F}_0^{ren}=\frac{Ak_BT}{\pi}\sum_{n,m=1}^\infty\frac{1}{m}\int_0^\infty dk_{||} k_{||} e^{-\widetilde{\beta}m\left(\frac{n^2\pi^2}{L^2}+k_{||}^2\right)^{1/2}}-V_pf^{(1)}_{bb}(T),
\end{equation}
which, under an appropriate change of variables, can be written as
\begin{equation}\label{37}
\Delta^{(1)}_T\mathcal{F}_0^{ren}=-\frac{Ak_BT\pi}{L^2}\sum_{n,m=1}^\infty\frac{n^2}{m}\int_1^\infty dx x e^{-\widetilde{\beta}mnx/L}-V_pf^{(1)}_{bb}(T).
\end{equation}
The integral can be written in terms of the modified Bessel's function of second kind \cite{abramowitz}, and we get
\begin{equation}\label{38}
\Delta^{(1)}_T\mathcal{F}_0^{ren}=-\frac{Ak_BT}{L^2}\sqrt{\frac{L}{\widetilde{\beta}}}\sum_{n,m=1}^\infty\left(\frac{n}{m}\right)^{3/2}K_{3/2}\left(\frac{\widetilde{\beta}mn\pi}{L}\right)-V_pf^{(1)}_{bb}(T).
\end{equation}
Thus, the total renormalized free energy is written as
\begin{eqnarray}\label{39}
\mathcal{F}_0^{(1),ren}&=&AL_p\overline{\epsilon}^{ren}_{vac}+\Delta^{(1)}_T\mathcal{F}_0^{ren}\nonumber\\
&=&-\frac{A\pi^2}{1440L_p^3}-C_1A(k_BT)^{3/2}L_p^{-3/2}\sum_{n,m=1}^\infty\left(\frac{n}{m}\right)^{3/2}K_{3/2}\left(\frac{\widetilde{\beta}mn\pi}{L}\right)+\nonumber\\
&+&C_2\frac{AL_p\pi^2(k_BT)^4}{90},\nonumber\\
\end{eqnarray}
where $C_1=\left(1+\frac{5M}{2R}-\frac{Ma\Omega}{R}\right)$ and $C_2=\left(1-\frac{6M}{R}+\frac{6Ma\Omega}{R}\right)$, are factors related with the gravitational field and the rotational parameters. $L_p$ is the proper distance between the plates.

Now, let us calculate the renormalized internal energy in the low and high temperature limits, which is given by
\begin{equation}\label{40}
U_0^{ren,1}(T)=-T^2\frac{\partial{(\mathcal{F}^{(1),ren}_0/T)}}{\partial T}.
\end{equation}
In order to obtain this quantity, we will take both low and high temperature limits of the total free energy (\ref{39}). For the first limit, we will consider the dominant term of the asymptotic expansion of the modified Bessel's function for large arguments \cite{abramowitz}, and thus we get
\begin{equation}\label{41}
U_0^{ren,1}(T)\thickapprox-\frac{A\pi^2}{1440L_p^3}+\frac{A}{\sqrt{2}}\left[C_3\pi L_p^{-3/2}(k_BT)+C_4L_p^{-5/2}(k_BT)^2\right]e^{-\widetilde{\beta}\pi/L}-C_2\frac{AL_p\pi^2(k_BT)^4}{30},
\end{equation}
where $C_3=\left(1+\frac{M}{2R}\right)$ and $C_4=\left(1+\frac{7M}{2R}-\frac{2Ma\Omega}{R}\right)$. Note that when $T\rightarrow0$, the renormalized internal energy coincides with the Casimir one at $T=0$.

The high temperature limit is obtained from the asymptotic expansion of the modified Bessel's function for small arguments \cite{abramowitz}. Thus we have
\begin{equation}\label{42}
U_0^{ren,1}(T)\thickapprox-\frac{A\pi^2}{1440L_p^3}-C_5\frac{A\zeta(3)(k_BT)^3}{\pi}-C_2\frac{AL_p\pi^2(k_BT)^4}{30},
\end{equation}
where $C_5=\left(1+\frac{M}{R}+\frac{2Ma\Omega}{R}\right)$. It is interesting to note that the second term in equation (\ref{42}) does not contribute to the thermal Casimir force $-\frac{\partial U_0^{ren}}{\partial L_p}$. The dominant term comes from the black-body radiation.

\subsection{Second order correction to the Casimir energy}

In the subsequent approximation, the black-body free energy density (\ref{34}) can also be readily obtained, and it is given by
\begin{equation}\label{43}
f^{(2)}_{bb}(T)=-\left(1+\frac{6L_pM}{R^2}-\frac{6Ma\Omega L_p}{R^2}\right)\frac{\pi^2(k_BT)^4}{90}.
\end{equation}
As in previous section, without gravity, Eq. (\ref{43}) represents the flat black-body free energy density. The total renormalized free energy is
\begin{eqnarray}\label{44}
\mathcal{F}_0^{(2),ren}&=&AL_p\overline{\epsilon}^{ren}_{vac,2}+\Delta^{(2)}_T\mathcal{F}_0^{ren}\nonumber\\
&=&-\frac{A\pi^2}{1440L_p^3}-C_6A(k_BT)^{3/2}L_p^{-3/2}\sum_{n,m=1}^\infty\left(\frac{n}{m}\right)^{3/2}K_{3/2}\left(\frac{\widetilde{\beta}mn\pi}{L}\right)+\nonumber\\
&+&C_7\frac{AL_p\pi^2(k_BT)^4}{90},\nonumber\\
\end{eqnarray}
where $C_6=\left(1-\frac{7ML_p}{4R^2}+\frac{ML_pa\Omega}{R^2}\right)$, $C_7=\left(1+\frac{6L_pM}{R^2}-\frac{6Ma\Omega L_p}{R^2}\right)$ and $L_p$ is the proper distance between the plates.

For this order of correction, let us calculate the renormalized internal energy in low and high temperature limits,  following the procedures adopted in the previous section, which is written as
\begin{equation}\label{45}
U_0^{ren,2}(T)=-T^2\frac{\partial{(\mathcal{F}^{(2),ren}_0/T)}}{\partial T}.
\end{equation}
In the low temperature limit, the renormalized internal energy is
\begin{equation}\label{46}
U_0^{ren,2}(T)\thickapprox-\frac{A\pi^2}{1440L_p^3}+\frac{A}{\sqrt{2}}\left[C_8\pi L_p^{-3/2}(k_BT)+C_9L_p^{-5/2}(k_BT)^2\right]e^{-\widetilde{\beta}\pi/L}-C_7\frac{AL_p\pi^2(k_BT)^4}{30},
\end{equation}
where $C_8=\left(1-\frac{3ML_p}{2R^2}\right)$ and $C_9=\left(1-\frac{5ML_p}{2R^2}+\frac{2ML_pa\Omega}{R^2}\right)$.

In the high temperature limit, we have
\begin{equation}\label{47}
U_0^{ren,2}(T)\thickapprox-\frac{A\pi^2}{1440L_p^3}-C_{10}\frac{A\zeta(3)(k_BT)^3}{\pi}-C_7\frac{AL_p\pi^2(k_BT)^4}{30},
\end{equation}
where $C_{10}=\left(1+\frac{ML_p}{2R^2}-\frac{2Ma\Omega L_p}{R^2}\right)$. We point out that the second term in equation (\ref{47}) does not also contribute to the thermal Casimir force $-\frac{\partial U_0^{ren,2}}{\partial L_p}$, and the dominant term is that of the black-body radiation.

\section{Final remarks}
We have calculated the Casimir energy for a massless scalar field in a region limited by two parallel uncharged conductor plates, in the presence of a slowly rotating spherical source ($a^2\simeq0$), in the weak field approximation regime. The plates are located tangentially to the surface, at the equatorial plane.

Initially, we considered the approximation in first order in terms of $M/R$ and introduced a non-inertial reference frame which has the same angular velocity of the sphere, in order to do the calculations. We showed that, like the static case analysed in \cite{sorge}, in which the Minkowski Casimir energy counterpart is not affected by the gravitational field in this order of approximation, when we consider that the source has angular momentum, we also do not get correction terms to the Casimir energy in flat Minkowski spacetime.

We also calculated the Casimir energy in the second order of approximation with respect to the expansion of the post-Newtonian parameter, which means that we considered the expansion up to $(M/R)^2$. It is worth noticing that in this case, there is an attenuation of the interaction between the plates that depend now on the parameters into play, including the surface gravity, spacetime rotation and non-inertial effects. The result found in \cite{sorge} is reobtained when the rotation is turned off.

Finally, we have examined the role played by the temperature in the Casimir effect, calculating the renormalized Helmholtz free energy and the internal energy of the system, in the low and high temperature limits, for both orders of approximation studied. We point out that the results obtained include naturally the thermal extension of the paper \cite{sorge}, when the rotation of the source is not considered. It is interesting to note that, despite the gravitational field does not manifest itself in the Casimir effect when we consider only terms $\mathcal{O}(M/R)$ at $T=0$, it shows up in the thermal corrections.

\section*{Acknowledgments}

V. B. Bezerra and H. F. Mota would like to thank CNPq for partial financial support. C.R. Muniz would like to thank to Universidade Federal da Para\'iba for the kind welcome and to CNPq for a Postdoctoral Fellowship.

\end{document}